\documentclass{aa}  
\usepackage{graphicx}
\usepackage{gensymb}
\usepackage{amsmath}
\usepackage{txfonts}
\usepackage{xcolor}

\begin{document}

   \title{Active regions and the large-scale magnetic field of solar cycle 24}
   
   \author{Ismo Tähtinen
                 \inst{1}
          \and
          Timo Asikainen\inst{1}
          \and
          Kalevi Mursula\inst{1}
          }
   \institute{Space Physics and Astronomy Research Unit, University of Oulu,
               POB 8000, FI-90014, Oulu, Finland\\
              \email{ismo.tahtinen@oulu.fi}
             }


 
  \abstract
   {Most of the intracyclic variability in the large-scale solar magnetic field comes from the equatorial dipole component of the solar magnetic field.
   The equatorial dipole component is highly sensitive to the longitude distribution of the active regions.
   }
   {We quantify the effect of individual active regions on the large-scale solar magnetic field of the solar cycle 24.
   We study the effect of the longitude distribution of active regions on the strength of the large-scale dipole component.
   }
   {We used a surface flux transport (SFT) model to simulate the evolution of individual active regions and quantified their effect on the large-scale magnetic field using the recently developed vector sum method.
   We took advantage of the longitudinal translational invariance of the SFT model and compared the observed solar cycle 24 to the 10~000 simulations of the solar cycle 24 using randomized longitudinal source locations, but otherwise identical flux emergence.
   }
   {We find that taking into account both the axial and equatorial components of the vector sum characterizing the global solar magnetic field sets better constraints on the parameter space of the SFT model than, for example, using the axial dipole moment alone as an optimization metric.
   We studied the maximum of cycle 24 and identified the recurrent and localized flux emergence in the southern hemisphere as the main culprit behind the rapid strengthening of the large-scale magnetic field in late 2014.
   We find that during the declining phase of the solar cycle, the strength of the large-scale magnetic field stayed above the median level of randomized simulations (p~$<$~0.027) for 42 subsequent rotations (from September 2014 to November 2017).
   This indicates that the longitudinal distribution of active regions is not random and,  rather, that it demonstrates a tendency for some regions to emerge at longitudes where their equatorial components reinforce the large-scale equatorial field.
   }
   {}


   \maketitle
%

\section{Introduction}\label{sec:Intro}
The magnetic field generated within the Sun’s convection zone can rise to the photosphere, where it manifests as active regions.
In the photosphere, their magnetic flux is spread across the solar surface through diffusion and advection which form the large-scale solar magnetic field.
The large-scale solar magnetic field plays an important role in shaping the near-Earth space environment because it is partly carried away by the solar wind to interplanetary space, where it forms the interplanetary magnetic field (IMF) enclosing all planetary bodies of the solar system.
The IMF plays a crucial role in shaping geophysical effects on Earth, acting as a direct link between the Sun and our planet, while also modulating the flux of cosmic rays throughout the solar system.

The evolution of active regions and their effect on the large-scale magnetic field was extensively studied by \citet{WangSheeley1991}, laying out the basic principles of active region evolution and their effect on the large-scale magnetic field.
The large-scale solar magnetic field is often characterized only by the dipole component, consisting of its axial and equatorial parts.
\citet{WangSheeley1991} showed that the axial dipole moment of individual active regions can either decay or grow, depending on the latitude and tilt angle of the active region.
Only active regions emerging at relatively low latitudes have the capacity to substantially affect the large-scale axial dipole component.
For the axial dipole contribution of an active region to persist, some of the leading-polarity flux must cross the equator against the opposing meridional flow, from where it can be transported to the opposite pole.
The evolution of the axial dipole moment of individual active regions and their effect on the large-scale axial dipole moment of the Sun have been the subject of great interest as the strength of the axial dipole at the solar minimum serves as a proxy for the amplitude of the following solar cycle \citep{Schatten1978,Svalgaard2005,Cliver2011,Upton2014,Cameron2016,Ijima2017,Jiang2018,Whitbread2018,Upton2018,Bhowmik2018,Yeates2020,Petrovay2020,Jiang2023}.

While the axial dipole component of active regions and large-scale solar magnetic field has been a subject of wide interest, the equatorial dipole has received less attention.
However, the large-scale equatorial dipole is important for the solar-terrestrial relations, as it modulates the strength of the IMF within intracycle timescales of $\tau~\sim~1$~yr \citep{WangSheeley2000a,WangSheeley2000b,WangSheeley2003}.
Unlike the axial dipole of active regions, which for low-latitude active regions can asymptotically approach a nonzero state where opposite polarities reside on opposite poles, the equatorial dipole always tends to zero on a timescale of $\tau~\sim~1$~yr.
This is because there is no mechanism that can sustain the longitudinal separation of opposite-polarity fluxes.
Although the equatorial dipole component of an individual active region decays quickly, its maximum strength is typically larger than that of the axial dipole component.
It  typically takes about six months for an active region to reach the maximum equatorial dipole strength \citep{WangSheeley1991,WangSheeley2003}.

The large-scale equatorial dipole component  crucially depends on the longitude distribution of the active regions, as it results from vectorially summing the individual equatorial components of all active regions.
The large-scale equatorial dipole can entirely cancel out even at the peak of the solar cycle if the equatorial dipoles of active regions are suitably out of phase with each other.
Conversely, if the equatorial dipole components of the active regions are suitably aligned, even relatively weaker regions can form a strong equatorial dipole.
The size distribution of active regions also matters, as fewer but larger active regions lead to larger fluctuations in the strength of the equatorial dipole because the larger unipolar regions diffuse more slowly than many active regions with the same flux \citep{WangSheeley2003}.

The most common quantity characterizing the global solar magnetic field is the so-called open solar flux (OSF), which is the total unsigned magnetic flux escaping the Sun.
The amount of open solar flux is related to the large-scale structure of the solar magnetic field.
The solar magnetic field is typically open only in large unipolar regions where the magnetic field lines can extend to high altitudes before turning back towards the surface.
Because the magnetic field lines from unipolar regions extend more freely, they can reach altitudes where they are swept away by the solar wind, while in more mixed regions the magnetic field closes back down at lower altitudes due to the proximity of the flux tube footpoints in the photosphere \citep{Cranmer2009,Wang2009}.
Coronal holes form in connection with large unipolar regions as the plasma can more easily escape along the open magnetic field lines.
Large unipolar regions and coronal holes are typically observed at polar regions during the solar minimum times, but they also form at lower latitudes \citep{Cranmer2009,Wang2009}.

Recently, \citet{Tahtinen2024} developed a vector sum method that quantifies the state of the large-scale solar magnetic field with a single vector.
This vector tracks the location of dominant unipolar magnetic fields and its magnitude closely matches with the OSF calculated from the potential field source surface (PFSS) model \citep{Altschuler1969,Schatten1969,Wang1992} with typical source surface radius of $2.5R_\odot$.
Vector sum can be naturally applied to study the axial and equatorial components of the solar magnetic field by projecting the vector sum parallel and perpendicular to the solar rotation axis.
In this paper, we coupled the vector sum method with the surface flux transport (SFT) model to study how the individual active regions affect the evolution of the large-scale solar magnetic field during solar cycle 24.
We focus on the effect that individual active regions and their longitude distribution have on the large-scale equatorial magnetic field.

In Sect.~\ref{sec2:Data}, we describe the data.
In Sect.~\ref{sec3:VectorSum}, we describe the vector sum method and apply it to synoptic HMI magnetograms.
In Sects.~\ref{sec4:SFT} and \ref{sec5:SFTOptimization}, we describe the SFT model and the optimization of the SFT parameters.
In Sect.~\ref{sec6:OSFPeak}, we combine the vector sum with the SFT model and analyze the evolution of individual active regions and their effect on the large-scale solar magnetic field during the rapid strengthening of the large-scale magnetic field in late 2014.
In Sect.~\ref{sec7:LongitudeDistribution}, we analyze the effect of the longitudinal distribution of active regions by performing multiple simulations of solar cycle 24 with randomized emergence longitudes.
We discuss our results in Sect.~\ref{sec:Discussion} and give our conclusions in Sect.~\ref{sec:Conclusions}.

\section{Data}\label{sec2:Data}
\subsection{Magnetograms}
In this paper, we use the pole-filled synoptic maps of the radial magnetic field (\texttt{hmi.synoptic\_mr\_polfil\_720s}) from the Helioseismic and Magnetic Imager on board the Solar Dynamics Observatory \citep[SDO/HMI;][]{Scherrer2012,Pesnell2012}.
Data covers Carrington rotations 2097--2297 (May 2010 to April 2025), but we mainly concentrate on rotations 2097--2224 (May 2010 to December 2019) that correspond to the solar cycle 24.
We have reduced the size of HMI maps to 180 pixels in sine of latitude and 360 pixels in longitude.

\subsection{Active regions}
We extracted active regions from HMI synoptic maps using an algorithm developed by \citet{Yeates2015}, which is available at GitHub.\footnote{\url{https://github.com/antyeates1983/sft_data}}
In brief, the algorithm smooths unsigned synoptic maps with a Gaussian filter ($\sigma=3$) and then finds continuous regions whose unsigned magnetic field strength exceeds a threshold $B_{\mathrm{par}}$.
We used a threshold of $B_{\mathrm{par}}=39.8$~G, which \citet{Whitbread2017} found to be optimal for SFT modeling in their parameter optimization study.
Finally, the algorithm corrects for flux balance by subtracting an equal amount from each pixel.
We extracted 672 active regions belonging to solar cycle 24 from 127 HMI synoptic maps.
Some of the extracted active regions are rather complex and consist of a few smaller nearby bipolar magnetic regions that the algorithm merges together.

The algorithm does not track the evolution of the magnetic field, so active regions may contain some residual magnetic flux from previous rotations.
For many applications, this is not a problem, but for our study it is important to avoid possible double counting of magnetic flux.
To correct for earlier magnetic flux, we used an SFT model to estimate the pre-existing magnetic flux inside the active region at the time of emergence.
We then corrected the active region magnetic field by subtracting the preceding magnetic field from the
active region.
This subtraction introduces a magnetic flux imbalance that was present in the preceding magnetic field, which we corrected by subtracting an equal amount of flux from each pixel so that the total magnetic flux remains zero.

We note that using the SFT model for flux correction is not entirely adequate for recurrent sunspots that persist longer than one Carrington rotation, as diffusion is reduced in strong sunspot fields.
This limitation, however, is a general challenge for classical SFT models, regardless of whether flux correction is applied.
Nevertheless, the correction we apply is equivalent to the flux adjustment implicitly performed in SFT models, where emerging active regions replace the existing magnetic field in the simulation \citep[e.g.,][]{Virtanen2017,Whitbread2018}.

\section{Vector sum and open solar flux}\label{sec3:VectorSum}
The vector sum is a straightforward method that, from a synoptic magnetogram, produces a single vector that describes the strength and orientation of large-scale solar magnetic field \citep{Tahtinen2024}.
The calculation of the vector sum starts by representing each pixel of a synoptic magnetogram as a vector in spherical coordinates.
Depending on the sign of the magnetic field, the vector either points outward (positive) or inward (negative) from the pixel, corresponding to the direction of the radial magnetic field.
The length of each vector is set to equal the total magnetic flux within the pixel, which also normalizes the length for different sized pixels.
The vector sum is simply the sum of these vectors, producing a single vector per magnetogram.
The magnitude of the vector sum $\Phi$ equals the total magnetic flux aligned with the vector sum axis (see Appendix~\ref{appendix}).

\begin{figure}[!htbp]
\resizebox{\hsize}{!}{\includegraphics{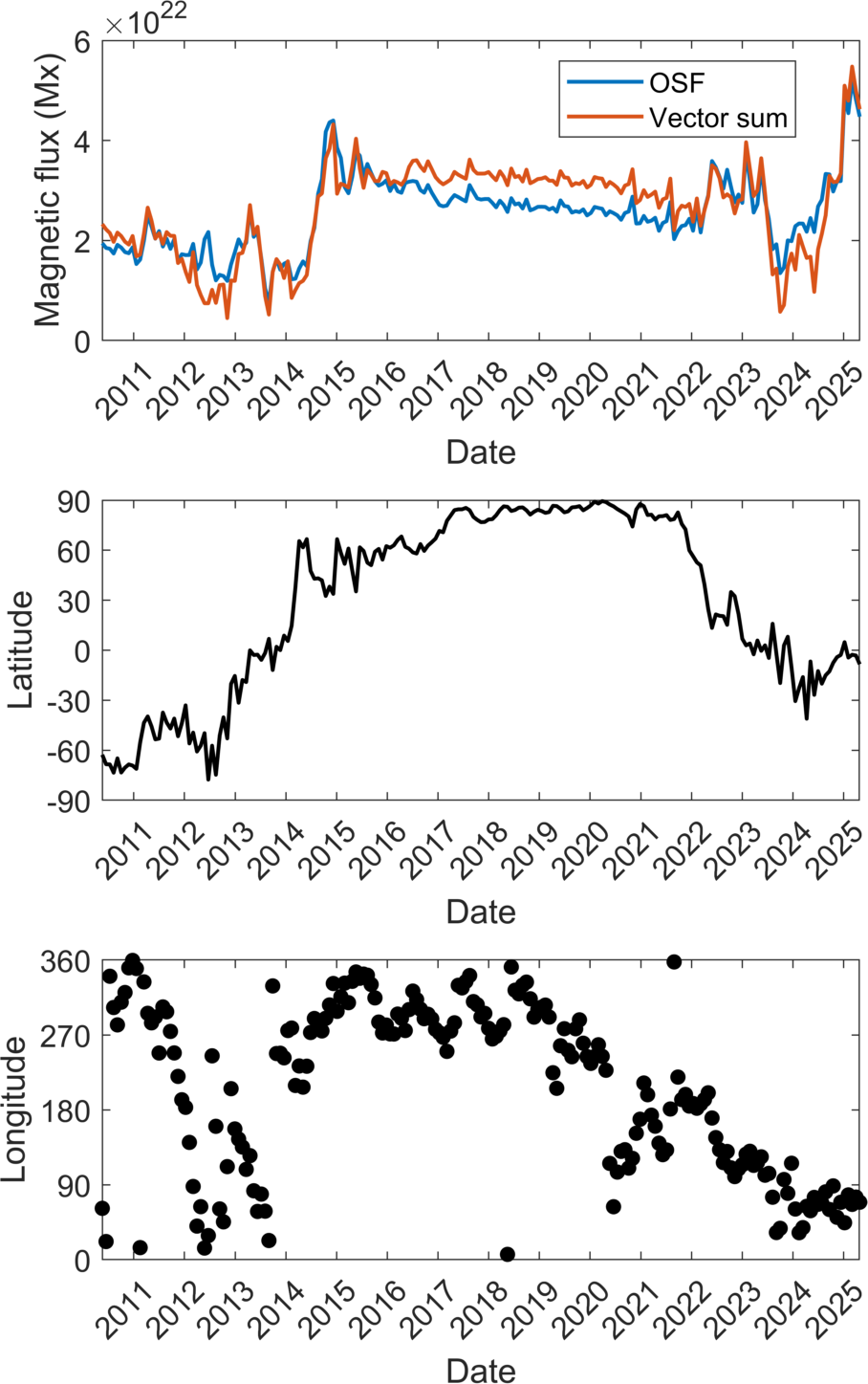}}
\caption{Vector sum magnitude, latitude and longitude. Upper panel: PFSS OSF ($l_{max}~=~50, R_{SS}=~2.5R_\odot$, blue) and the absolute value of the vector sum (orange) from the HMI synoptic magnetograms. Middle panel: Vector sum latitude. Lower panel: Vector sum longitude.
}\label{fig:VectorSumAndOSF}
\end{figure}

The magnitude of the vector sum $\Phi$ closely matches with the open solar flux calculated from the PFSS model \citep[$l_{max}~=~50, R_{SS}=~2.5R_\odot$,][]{Tahtinen2024}.
This is demonstrated in Fig.~\ref{fig:VectorSumAndOSF}, which shows the vector sum magnitude together with the PFSS OSF calculated from the HMI synoptic magnetograms for the past 15 years.
The vector sum is closest to the PFSS around the solar maximum in 2013-2015 and shows the largest difference towards the end of the cycle and around the solar minimum in 2016--2021.

Figure \ref{fig:VectorSumAndOSF}b shows the evolution of the vector sum latitude $\lambda$ for HMI data.
Vector sum is directed toward the south at the start of solar cycle 24 and toward the north at the start of solar cycle 25, that is, always toward dominantly positive polarity regions.
Using magnetograms from 1976 onward \citet{Tahtinen2024} show that the vector sum latitude follows the Hale cycle, traveling from pole to pole during one solar cycle.
The vector sum latitude can be used to estimate the tilt angle of the heliospheric current sheet through the vector sum co-latitude, $90\degree{}-|\lambda|$. 

Figure \ref{fig:VectorSumAndOSF}c shows the vector sum longitude for HMI data.
Vector sum longitude displays long periods of rather consistent evolution in the Carrington frame.
For example, the vector sum longitude stayed within about 90\degree{} wide band (270\degree{}-360\degree{}) from the late 2014 until early 2019.
\citet{Tahtinen2024} find that the direction of the vector sum often corresponds to the location of coronal holes and that their apparent motion in the Carrington frame is similar to the motion of low-latitude coronal holes \citep{Krista2018,Harris2022}.

\begin{figure*}[!htbp]
\resizebox{\hsize}{!}{\includegraphics{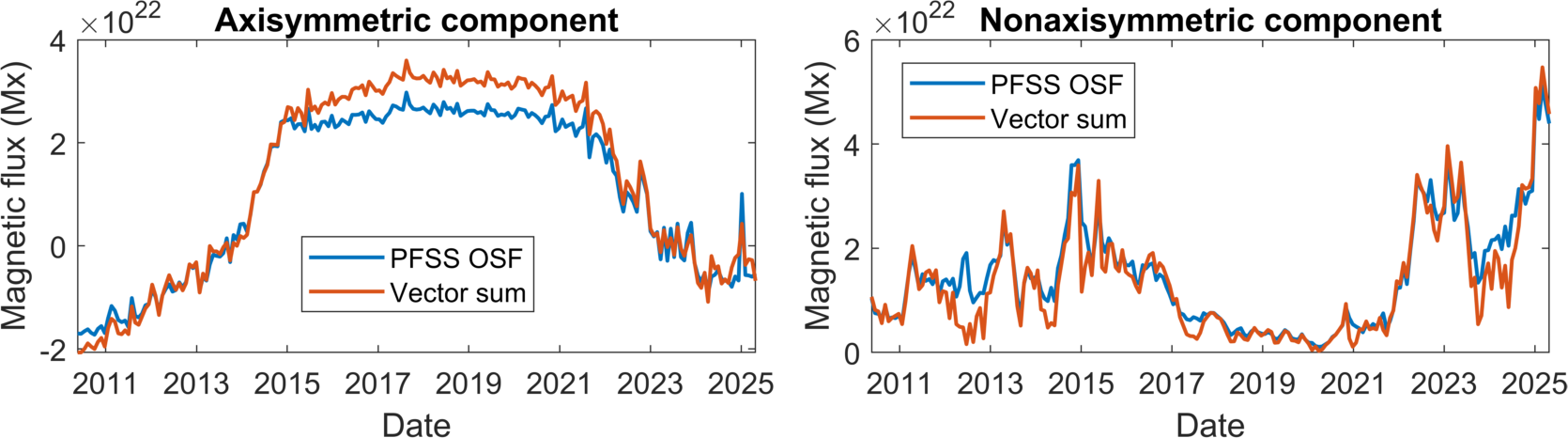}}
\caption{Axisymmetric (left panel) and nonaxisymmetric (right panel) component of the PFSS OSF ($l_{max}~=~50, R_{SS}=~2.5R_\odot$, blue) and the vector sum (orange). The vector sum components correspond to projections of the vector sum along ($\Phi\sin{\lambda}$, axial) and perpendicular ($\Phi\cos{\lambda}$, equatorial) to solar rotation axis. The PFSS OSF components correspond to PFSS expansion using only axisymmetric (m=0) and nonaxisymmetric components ($m\neq0$).
}\label{fig:VectorSumAndOSFProjections}
\end{figure*}

The left panel of Fig.~\ref{fig:VectorSumAndOSFProjections} shows the axial ($\Phi\sin{\lambda}$) component of the vector sum (orange) and the axisymmetric ($m=0$) part of the PFSS OSF (blue).
We have multiplied the axisymmetric OSF, which is unsigned value, with the sign of the axial vector sum to take into account the polarity change.
The axial component of the vector sum and the axisymmetric OSF evolve rather steadily.
At the start of solar cycle 24, the axial components are negative and weaken until the end of October 2013, when the axial component changes sign and begins to rapidly strengthen shortly thereafter.
The rapidly increasing phase lasts about a year, culminating at the start of 2015 when the total vector sum and the PFSS OSF reach their maximum value in solar cycle 24.
Figure~\ref{fig:VectorSumAndOSFProjections}a also reveals that the difference between the vector sum and the PFSS OSF in Fig.~\ref{fig:VectorSumAndOSF} is mainly due to the difference in axial components.
The difference is greatest close to solar minimum because during these times the structure of the solar magnetic field is closest to a simple dipole field.
In Appendix~\ref{appendix} we show that the magnitude of the vector sum equals the PFSS OSF, when only the dipole component (l~=~1) is included and the source surface radius is set to $R_{ss}\approx2.14R_\odot$.
Because the vector sum corresponds to the dipole term with $R_{ss}=2.14R_\odot$, it necessarily becomes larger than PFSS OSF with $R_{ss}=2.5R_\odot$ during the times that the solar magnetic field resembles a pure dipole (PFSS OSF generally decreases with increasing $R_{ss}$).
Appendix~\ref{appendix} also shows that the axial and equatorial components of the vector sum are proportional to the axial and equatorial dipole moments of the Sun.
Although the vector sum is directly related to the solar dipole moment, it should be kept distinct because it has units of flux rather than flux density and because it directly relates PFSS OSF to the photospheric flux distribution.

Figure~\ref{fig:VectorSumAndOSFProjections}b shows the nonaxisymmetric components of the vector sum (orange, $\Phi\cos{\lambda}$) and the PFSS OSF (blue, $m\neq0$).
Although the axial component of the vector sum naturally picks up a sign from its direction with respect to the solar rotation axis, the equatorial component is always positive since it represents the magnitude of the vector projected onto the equatorial plane. 
Figure~\ref{fig:VectorSumAndOSFProjections}b shows that the nonaxisymmetric components of the vector sum and the PFSS OSF agree well throughout the studied period except for some shorter periods when the nonaxisymmetric PFSS OSF increases relative to the equatorial vector sum, as most clearly seen in 2012 and 2024.
These differences are also visible in panel a of Fig.~\ref{fig:VectorSumAndOSF}.
Such behavior indicates that the nonaxisymmetric structure of the solar magnetic field became more complex during these times, and higher-order multipoles, which the vector sum does not capture, contributed substantially to the strength of the nonaxisymmetric magnetic field.
The evolution of the equatorial component of the vector sum is much more variable than that of the axial component, showing large fluctuations with separate peaks in 2011, 2013, 2014, 2022, and 2025.
These peaks are also visible in Fig.~\ref{fig:VectorSumAndOSF}a.

\section{Surface flux transport model}\label{sec4:SFT}
We used a surface flux transport (SFT) model to study the evolution of the large-scale solar magnetic field (see, e.g., Yeates 2023 for review).
The SFT model describes the evolution of the radial magnetic field subject to differential rotation $\Omega(\theta)$, meridional flow $u_\theta(\theta)$, and turbulent diffusion $\eta$ with the induction equation
\begin{equation}
\frac{\partial B_r}{\partial t} + \nabla_h \cdot (\mathbf{u}_h B_r) = \eta \nabla_h^2 B_r + S,
\end{equation}
where $\mathbf{u}_h$ describes the horizontal flow velocity (due to $\Omega(\theta)$ and $u_\theta(\theta)$) and the magnetic flux emergence is modeled with the source term S.
The explicit form of the governing equation in spherical coordinates is
\begin{align}
    \frac{\partial B_r}{\partial t} = &-\frac{1}{R_{\odot} \sin \theta} \frac{\partial}{\partial \theta} \left( \sin \theta \, u_{\theta} B_r \right)
    - \Omega(\theta) \frac{\partial B_r}{\partial \phi}\\
    &+ \frac{\eta}{R_{\odot}^2 \sin \theta} \frac{\partial}{\partial \theta}  
    \left( \sin \theta \frac{\partial B_r}{\partial \theta} \right)
    + \frac{\eta}{R_{\odot}^2 \sin^2 \theta} \frac{\partial^2 B_r}{\partial \phi^2} + S.
\end{align}
Here, we use the differential rotation profile of \citet{Snodgrass1990}:
\begin{equation}
    \Omega(\theta) = 0.18 - 2.396 \cos^2(\theta) - 1.787 \cos^4(\theta) \quad [^{\circ} \text{day}^{-1}],
\end{equation}
and the meridional flow profile of \citet{Whitbread2018}:
\begin{equation}\label{eq:MeridionalFlow}
        u_{\theta}(\theta) = -u_0 \sin^p{\theta}\cos{\theta},
\end{equation}
with their shape parameter $p = 2.33$.
Diffusivity, $\eta$, and meridional flow amplitude, $u_0$, are left as the free parameters of the model.
We describe the optimization of these parameters in Sect.~\ref{sec5:SFTOptimization}.

The details of the used SFT model are described in \citet{Virtanen2017}, except for some modifications discussed below.
Firstly, unlike \citet{Virtanen2017}, who used the SFT model to simulate the evolution of the solar magnetic field over many cycles, we did not use the radial decay term in our simulations since we found it unnecessary for the evolution of the large-scale magnetic field over the much shorter period studied here.
Secondly, we modified the way the active regions are inserted into the simulation.
In \citet{Virtanen2017}, new active regions replaced the existing pixels when they were inserted in the simulation, which made the model nonlinear because the magnetic flux in the simulation needs to be balanced by adding the flux imbalance of the replaced pixels back to the simulation.
If the new active region flux is added on top of the existing field instead of replacing it, the SFT model becomes linear.
However, there is a drawback that some of the magnetic flux from previous rotations could be inserted into the model twice, since the active regions extracted from the synoptic magnetograms include some magnetic flux remnants of previous active regions.
As discussed in Sect.~\ref{sec2:Data}, we corrected the active region data for preceding flux in order to avoid inserting the same flux twice into the simulation.

A linear SFT simulation using the flux-corrected dataset produces results equivalent to a nonlinear simulation based on the original uncorrected active region data.
However, the current approach allows us to better simulate the evolution of each active region independently, enabling us to quantify and track the individual contributions of active regions to the large-scale solar magnetic field over time.

\section{Parameter optimization and response of vector sum components}\label{sec5:SFTOptimization}
\begin{figure*}[!htbp]
\resizebox{\hsize}{!}{\includegraphics{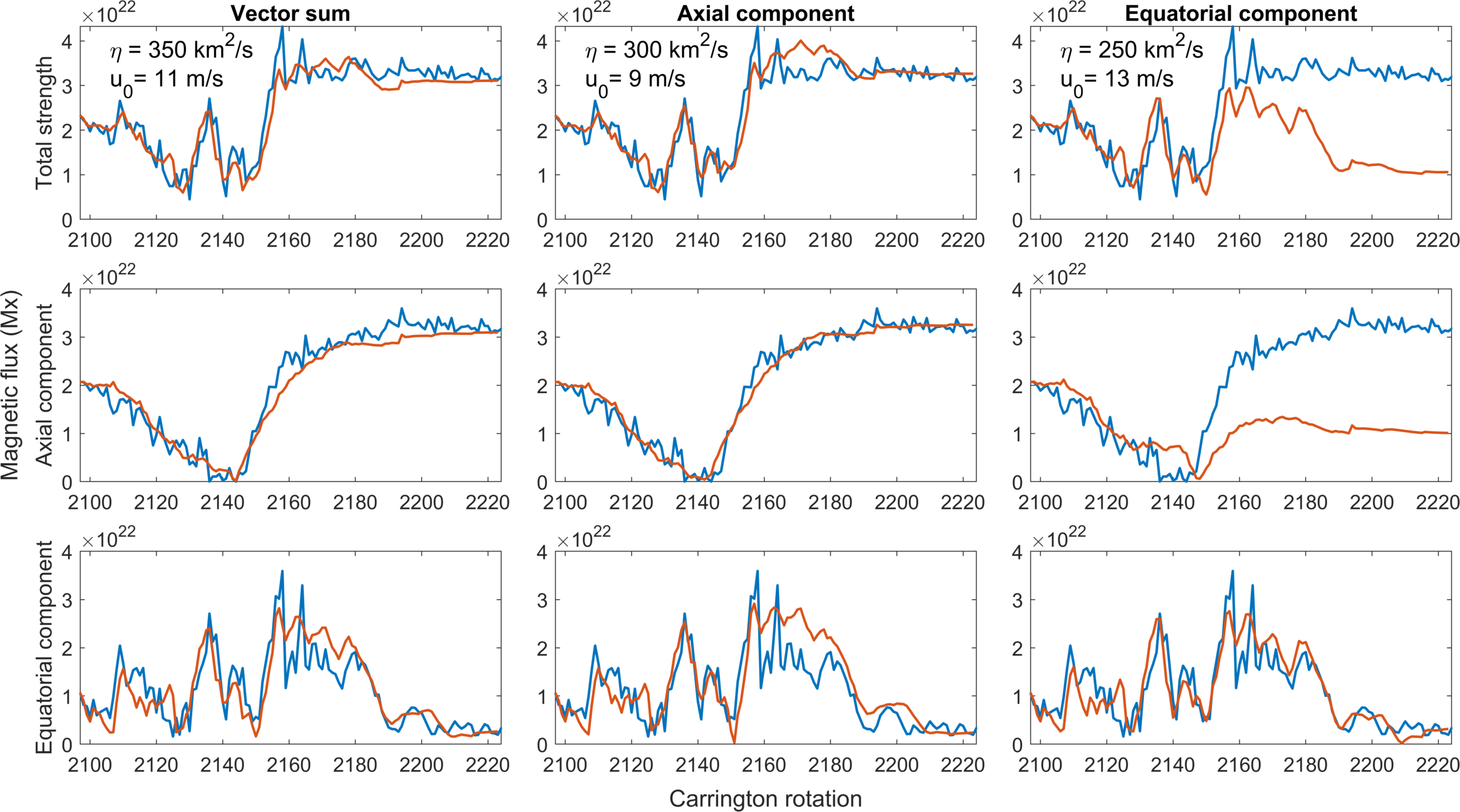}}
\caption{Best-fit simulations. Left column:  Best-fit simulations based on the total vector sum. Middle column: The best-fit simulations based on the axial component. Right column:  Best-fit simulations based on the equatorial component. First row: Total vector sum. Middle row: Axial component. Lower row: Equatorial component. Blue color shows the vector sum from HMI magnetograms and orange from the SFT simulations. 
}\label{fig:BestSFTParameters}
\end{figure*}

The free parameters, diffusivity, $\eta$, and meridional flow amplitude, $u_0$, need to be tuned for the SFT model.
We fitted these parameters by running the SFT model with different values, with $\eta$ ranging from 200~$\mathrm{km^2/s}$ to 600~$\mathrm{km^2/s}$ in steps of 50~$\mathrm{km^2/s}$, and $u_0$ ranging from 9 m/s to 20 m/s in steps of 1 m/s.
We used the synoptic map of Carrington rotation (CR) 2097 (May 2010) as an initial state and simulated the evolution of the photospheric magnetic field until the end of the solar cycle 24 (CR 2224, November 2019).
We optimized the parameters by comparing the vector sum of the simulated synoptic maps to the vector sum calculated from the HMI synoptic maps.
The best fit based on the root mean square error (RMSE) between the simulations and the HMI observations was found with parameters $\eta~=~350~\mathrm{km^2/s}$ and $u_0~=~11$~m/s.
The vector sum from the simulation is shown in the upper panel of left column in Fig.~\ref{fig:BestSFTParameters} together with the vector sum from the HMI maps.
The panels below show the axial (middle) and equatorial components (lower) for simulations and HMI observations obtained from the same optimization.

We also calculated the best-fit parameters after optimizing the axial (middle column) and the equatorial (right column) components of the vector sum, respectively.
Figure~\ref{fig:BestSFTParameters} shows that there is a trade-off between the best-fit parameters for the axial and equatorial components.
The best-fit parameters ($\eta~=~300~\mathrm{km^2/s}$ and $u_0~=~9$~m/s) for the axial components lead to a too strong equatorial component, while the best-fit parameters ($\eta~=~250~\mathrm{km^2/s}$ and $u_0~=~13$~m/s) for the equatorial component lead to an overly weak axial component and total flux.

\begin{figure*}[!htbp]
\resizebox{\hsize}{!}{\includegraphics{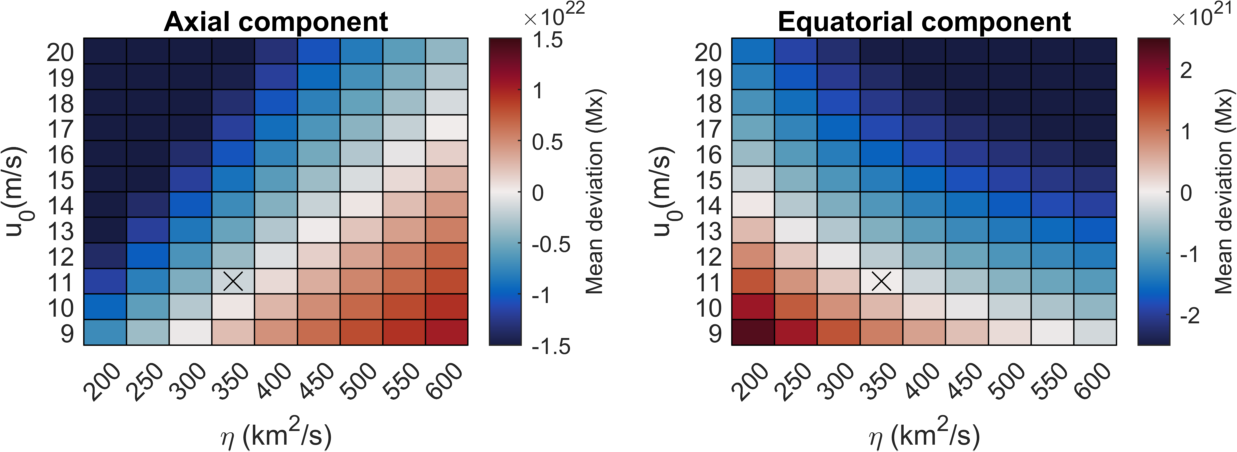}}
\caption{Effect of SFT parameters on the axial and equatorial components. Left panel: Axial component. Right panel: Equatorial component. Color shows the mean deviation between the SFT simulation and HMI vector sum components. The best-fit parameters $\eta~=~350~\mathrm{km^2/s}$ and $u_0~=~11$~m/s obtained using the total vector sum is marked with a cross. 
}\label{fig:AxialAndEquatorialParameters}
\end{figure*}

The conflicting situation between the axial and the equatorial components is depicted in Fig.~\ref{fig:AxialAndEquatorialParameters}, which shows the mean deviation of the axial (left panel) and equatorial (right panel) components for each of the simulations as a function of $\eta$ (x-axis) and $u_0$ (y-axis).
The left panel shows that there is a rising diagonal-shaped parameter range (white pixels) of ($\eta,u_0$) values from (300,9) to (600,18) that produces close to equally well-matching fits between the axial component of simulation and observations.
Contrastingly, the right panel of Fig.~\ref{fig:AxialAndEquatorialParameters} shows that there is an analogous descending diagonal of ($\eta,u_0$) values from (200,14) to (500,9) that produces equally good solutions for the equatorial component.
The best-fit parameters $\eta~=~350~\mathrm{km^2/s}$ and $u_0~=~11$~m/s (marked by X in both panels) that are based on the full vector sum lie in the cross-section of these opposing trends depicted by the axial and equatorial components.

\section{Evolution of individual active regions}\label{sec6:OSFPeak}
\subsection{Quantifying the influence of active regions}
Since vector summing is linear, we can apply it piece by piece to any linear combination of the original map and then add the partial sum vectors together.
This means that we can separate the contribution of each active region on the large-scale magnetic field as characterized by the vector sum.
Furthermore, the SFT model allows us to track how the dispersing flux from each active region individually influences the large-scale magnetic field after emergence.

We simulated the individual evolution of each active region in our dataset using the SFT model.
For each individual simulation, we calculated the vector sum at the start of every Carrington rotation up to CR2224.
This produces an ensemble of vector timeseries, where each series represents the evolution of the large-scale dipole component of a single active region.
Due to the linearity of the vector sum and the SFT model, it is enough to work just with these vector series when studying the effect of the active regions on the large-scale magnetic field (with the vector sum).
Summing these vector sum timeseries together produces a series equivalent to the vector sum calculated from the full SFT simulation.
From now on, we  refer to the vector sum of individual active regions as active region vectors and the vector sum over the full Sun as a large-scale vector.

The effect of each active region can be quantified, for example, by projecting the active region vector along the large-scale field vector, which represents a contribution of particular active region on the large-scale field.
Alternatively, we can leave one active region vector out when summing vectors together to produce the large-scale field vector and calculate how this affects the large-scale field. 
As the effect of the active regions on the large-scale axial component has been much more thoroughly studied  \citep[e.g.,][]{Jiang2014,Nagy2017,Jiang2018, Petrovay2020}, we focus on the effect that active regions have on the equatorial component of the large-scale field.

\subsection{Peak of solar cycle 24}
The strength of the large-scale solar magnetic field increased rapidly within a half-year period in the late 2014 and early 2015 (see Fig.~\ref{fig:VectorSumAndOSF}).
Since we  simulated the evolution of each active region separately, we were able to calculate which active regions had the largest effect on the large-scale solar magnetic field during this increase.
Here, we analyze how the active regions that emerged between Carrington rotations 2140 (August 2013) and 2157 (November 2014) affected the large-scale solar magnetic field until the end of CR2157 when the equatorial component of the large-scale solar magnetic field reached its peak strength (see Fig.~\ref{fig:VectorSumAndOSFProjections}b).
 
\begin{figure*}[!htbp]
\resizebox{\hsize}{!}{\includegraphics{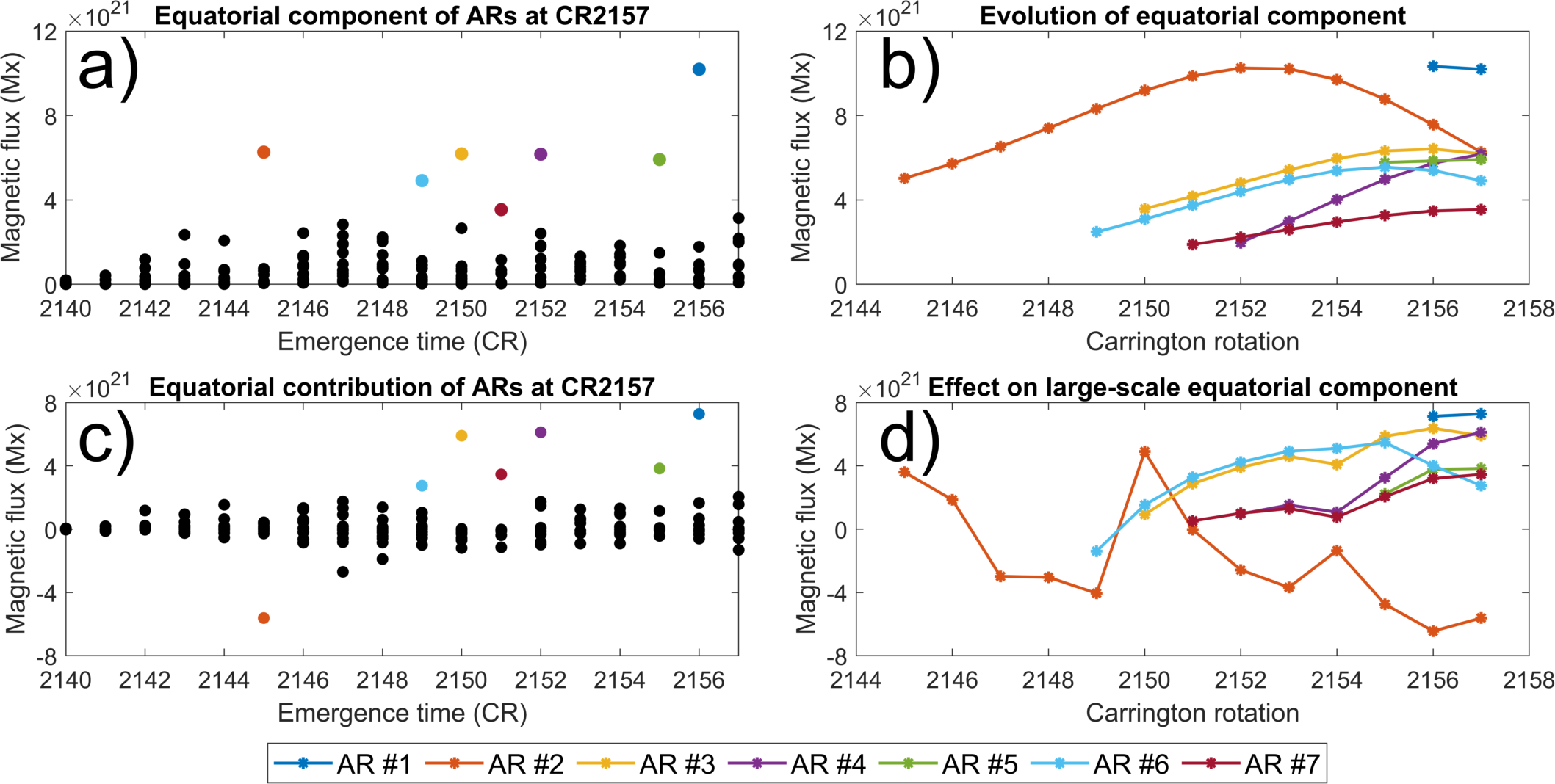}}
\caption{Equatorial components of active regions. (a) Equatorial component at CR2157 vs. time of emergence. (b)  Evolution of equatorial components of seven active regions which had strongest equatorial component at CR2157. (c)  Projections of the active region equatorial components along the large-scale field at CR2157. (d) Evolution of contributions of seven active regions which had strongest equatorial component at CR2157. Colors mark the seven active regions that had the strongest equatorial component at CR2157 and correspond to same active regions in all panels.
}\label{fig:OSFPeakAREvolution}
\end{figure*}

Figure~\ref{fig:OSFPeakAREvolution}a shows the strength of the equatorial component of active regions at the end of CR2157 plotted against the Carrington rotation of their emergence.
We  colored the markers of the seven active regions with the strongest equatorial components at the end of CR2157 and label these regions AR\#1 to AR\#7, with AR\#1 being the strongest and AR\#7 the weakest out of the seven.
Figure~\ref{fig:OSFPeakAREvolution}a shows that most of the active regions with the strongest equatorial components emerged already many rotations earlier.
AR\#2 with the second strongest equatorial component emerged almost a year (12 rotations) before CR2157.
AR\#1 with the strongest equatorial component, emerged during the previous CR2156 and corresponds to NOAA region 12192, which was the largest active region observed in 24 years.

Figure~\ref{fig:OSFPeakAREvolution}b shows the evolution of the equatorial component of seven active regions colored in Fig.~\ref{fig:OSFPeakAREvolution}a.
The equatorial component of all but AR\#1 increases first with time, reaching their peak strength about six months after their emergence.
Such behavior is typical and agrees with the results of \citet{WangSheeley1991,WangSheeley2003}.
The initial equatorial component of these regions increases to the maximum by a factor of about 2-3.
The peak equatorial strength of AR\#2 was at the same level as the peak equatorial strength of AR\#1, although the initial strength of the latter was twice that of the former.
Unlike other active regions, the equatorial component of AR\#1 obtained a maximum high value immediately after its emergence.
This is due to the almost zero tilt angle that this region had (Tähtinen et al., in prep.).
Due to the small tilt angle, there is no latitudinal separation between the opposite polarities, which is necessary for shearing by differential rotation to amplify the equatorial component.

Figure~\ref{fig:OSFPeakAREvolution}c is similar to Figure~\ref{fig:OSFPeakAREvolution}a, but instead of the strength of the equatorial component it shows the contribution of active regions to the equatorial component of the large-scale field.
We quantified the active region contribution by projecting the active region vector along the equatorial component of the large-scale vector.
The coloring of the markers is the same as in Fig.~\ref{fig:OSFPeakAREvolution}a.
Because the active region vectors can point to any longitudinal direction in the equatorial plane, they can either enhance (positive) or decrease (negative) the overall equatorial component of the large-scale magnetic field.
Figure~\ref{fig:OSFPeakAREvolution}c shows that at CR2157 six out of seven of these active regions had a large positive effect on the large-scale equatorial component, while AR\#2 made a rather large negative contribution.

Figure~\ref{fig:OSFPeakAREvolution}d shows how the contribution of these active regions evolved before CR2157. 
Although the contribution of AR\#2 during the CR2157 was negative, at the time of its emergence and even in the middle of its evolution it was contributing positively to the large-scale magnetic field.
Conversely, AR\#6 was contributing negatively at the time of its emergence at CR2149 before turning positive at CR2150.
The contribution of AR\#2 briefly turned positive at the same time before turning back to negative.
After CR2150, the contribution of AR\#2 became increasingly negative, while the other active regions show an increasing positive trend in their contributions.

\begin{figure*}[!htbp]
\resizebox{\hsize}{!}{\includegraphics{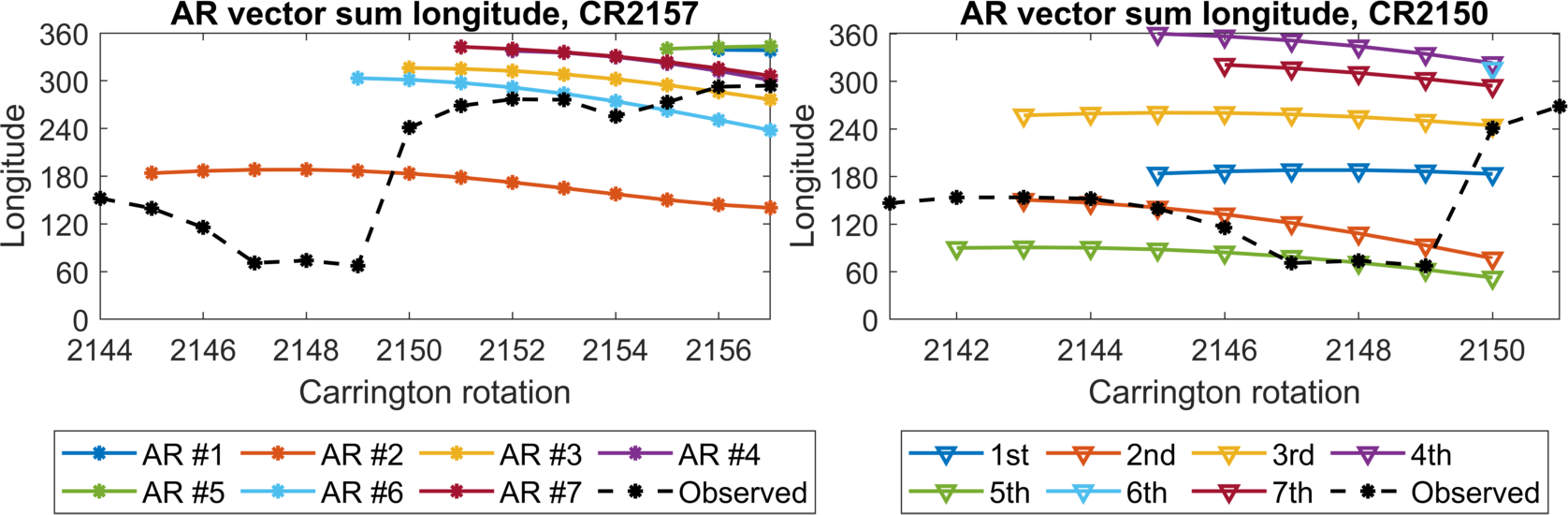}}
\caption{Evolution of active regions vector sum longitude. Left: Active regions that had strongest equatorial component at CR2157. Colors mark the same regions as in Fig.~\ref{fig:OSFPeakAREvolution}. Right: Active regions that had strongest equatorial component at CR2150. The dashed black line shows the location of vector sum longitude of the large-scale field. 
}\label{fig:ARDipoleLongitude}
\end{figure*}

The reason for the different behavior of AR\#2 from the rest of the active regions can be seen from the left panel of Fig.~\ref{fig:ARDipoleLongitude}, which shows how the vector sum longitude of these active regions evolved during the same time period.
While the vector sum longitude of AR\#2 starts at about 180\degree{} and drifts to 140\degree{} at CR2157, the vector sum longitudes of other active regions stay between 240\degree{} and 360\degree{}.
Although the evolution of vector sum longitude of active regions is smooth and gradual, the vector sum longitude of the large-scale field (black) goes through an abrupt 180\degree{} jump from longitude 67\degree{} to 241\degree{} between CR2149 and CR2150.
At the time of its emergence the vector sum longitude of AR\#2 is rather close to that of the large-scale field at 140\degree{} and thus the contribution of AR\#2 is positive.
As the vector sum longitude of the large-scale field drifts toward smaller longitudes, the distance between the active region and the large-scale field vector increases, and the contribution from AR\#2 becomes negative.
At CR2150, when the large-scale field makes an abrupt 180\degree{} flip, the distance between the AR\#2 and large-scale field briefly decreases, causing the positive spike in Fig.~\ref{fig:OSFPeakAREvolution}d.
After CR2150 the direction of the large-scale vector drifts in the direction of the other six active region vectors.
The abrupt 180\degree{} flip of the large-scale vector sum longitude indicates that magnetic flux from active regions with oppositely directed equatorial component forced the large-scale field to align with this new flux configuration.
The equatorial component of the large-scale field can be seen to reach a minimum value around CR2150 in the lower left panel of Fig.~\ref{fig:BestSFTParameters}.

The right panel of Fig.~\ref{fig:ARDipoleLongitude} shows the evolution of the vector sum longitude for active regions whose equatorial components were the largest at CR2150.
It shows that there was competition between active regions whose equatorial components were aligned with the direction of the large-scale field before and after CR2150.
It also shows that the seed for the rapid increase of the large-scale equatorial component was already placed well before CR2150.

\begin{figure}[!htbp]
\centering
\includegraphics[width=\textwidth,height=0.95\textheight,keepaspectratio]{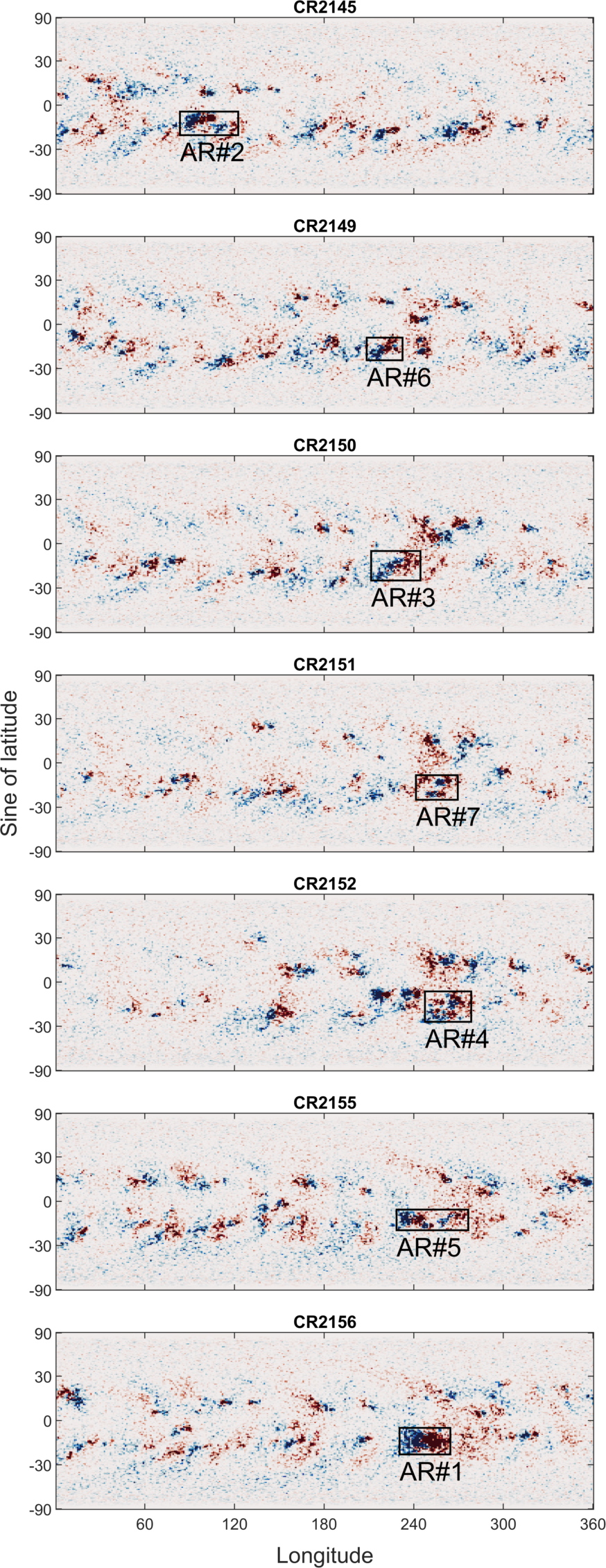}
\caption{Locations of the seven active regions that had strongest equatorial component at CR2157 at the time of their emergence.
}\label{fig:Magnetograms}
\end{figure}

Figure~\ref{fig:Magnetograms} shows the locations of the seven active regions on synoptic maps at the time of their emergence.
While all other active regions emerged fairly close to longitude 240\degree{}, AR\#2 emerged on the same hemisphere but almost on the opposite longitude.
Because these seven regions obeyed Hale's law, the equatorial vectors of six active regions around longitude 240\degree{} were also directed in the opposite direction compared to AR\#2.
The repeated flux emergence in the southern hemisphere around Carrington longitude 240\degree{} first decreased the existing equatorial field and then flipped the large-scale magnetic field in the opposite direction.

\section{Effect of longitude distribution}\label{sec7:LongitudeDistribution}
\begin{figure*}[!htbp]
\resizebox{\hsize}{!}{\includegraphics{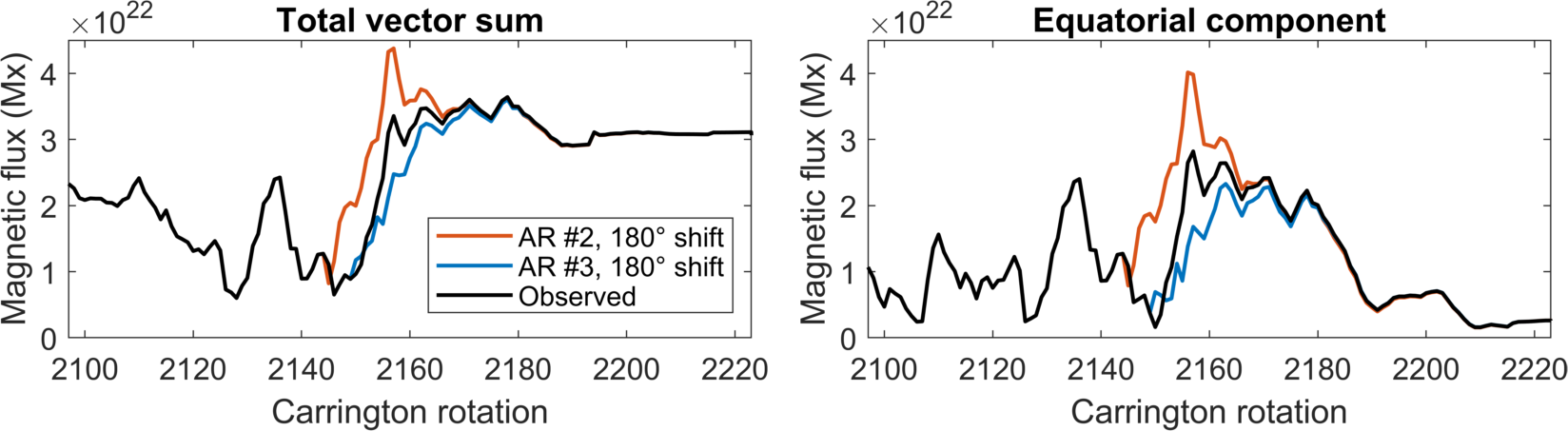}}
\caption{Effect of shifting active regions \#2 (orange) and \#3 (blue) 180\degree{} in longitude. Left:\ Total vector sum. Right: Equatorial component. Black line shows the unaltered SFT simulation. 
}\label{fig:180LongitudeShift}
\end{figure*}

As the emergence longitude of an active region largely determines the direction of its equatorial component, the longitude distribution of active regions is crucial in determining the strength of the large-scale equatorial component which collectively arises from the individual active regions.
We further demonstrate the effect of emergence longitude in Fig.~\ref{fig:180LongitudeShift}, which shows how shifting the emergence longitudes of AR\#2 and AR\#3  by 180\degree{} changes the strength the large-scale field (left panel) and its equatorial component (right panel).
We performed the calculation by adding 180\degree{} to longitude component of these two active region vectors and then summing over all active region vectors.
Due to the invariance of the SFT model under longitude translations, this calculation is equivalent to performing separate SFT simulation with the same shift introduced in emergence location of the active region.
The complete simulation from CR2097 (May 2010) to CR2224 (November 2019) is presented to illustrate the effect of these regions on the large-scale field over nearly a full solar cycle.
The longitude shift has no effect on the axial component as there is no change in latitudinal location of magnetic flux.

Shifting AR\#2 by 180\degree{} in longitude increases the maximum strength of the large-scale magnetic field at CR2157 by $1.0\times10^{22}$~Mx (30\%), while shifting the AR\#3 by 180\degree{} decreases it by $0.9\times10^{22}$~Mx (26\%).
For the equatorial component, shifting AR\#2 by 180\degree{} longitude increases the maximum strength at CR2157 by $1.2\times10^{22}$~Mx (41\%), while shifting the AR\#3 by 180\degree{} decreases it by $1.1\times10^{22}$~Mx (40\%).
The equatorial component also starts to increase five rotations earlier as the AR\#2 starts to contribute constructively to the large-scale field from CR2146 onwards instead of suppressing it.
While both active regions have a rather larger effect on the large-scale magnetic field, the difference between the original and altered simulations disappear within two years as the equatorial components of the active regions die off.

\begin{figure*}[!htbp]
\resizebox{\hsize}{!}{\includegraphics{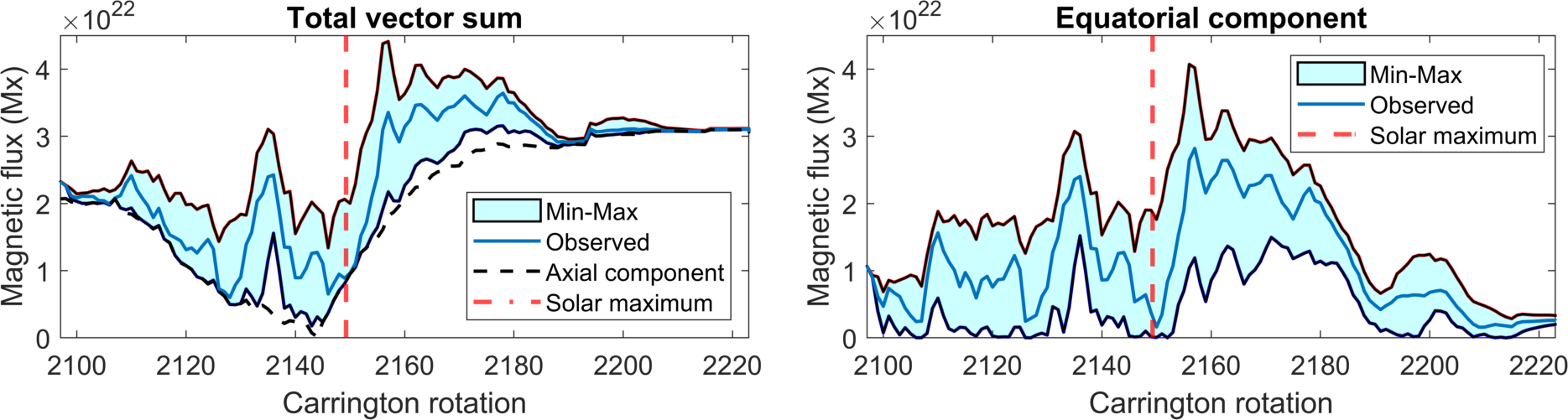}}
\caption{Minimum and maximum vector sum resulting from shifting a single active region in longitude. Left: Total vector sum. Right: Equatorial component of the vector sum.  The shaded region shows the minimum-maximum range. The solid blue line shows the unaltered simulation. The red dashed line shows the location of the solar maximum (CR2149, April 2014, SILSO 13-month smoothed mean). The  dashed black line in the left panel shows the strength of the axial component.
}\label{fig:PeakFluxSingleAR}
\end{figure*}

As shown in Fig.~\ref{fig:180LongitudeShift}, the emergence longitude of even a single active region can have a considerable effect on the large-scale magnetic field.
However, the effect of one active region is not always immediately obvious, as it depends on the factor by which the equatorial component grows over time, and on the locations of the other active regions.
In Fig.~\ref{fig:PeakFluxSingleAR}, we show the minimum-maximum range for the effect that a change in the longitude of a single active region causes for the total vector sum (left panel) and its equatorial component (right panel).
We constructed this plot by shifting the emergence longitude of each active region (at a time) in increments of 1\degree{} from 1\degree{} to 360\degree{} and recorded the minimum and maximum large-scale vector sum for each Carrington rotation.
We note that vector sum presentation makes this calculation much more feasible since, in principle, the calculation would amount to performing 360*672=241920 SFT simulations.
We also show the strength of the axial component in Fig.~\ref{fig:PeakFluxSingleAR}, since it equals the lower boundary of the large-scale vector sum when the strength of the equatorial component is zero.

The minimum range closely follows the absolute minimum given by the strength of the axial component in the left panel of Fig.~\ref{fig:PeakFluxSingleAR} which corresponds to values close to zero in the right panel.
This indicates that for large parts of solar cycle 24, it was possible to cancel the equatorial component of the large-scale field just by shifting the longitude of a single active region.
However, during 2013-2014 and from mid-2014 to 2017, the minimum range remained clearly above zero, showing that during these times, it was not possible to cancel the equatorial component by just shifting the longitude of one active region.
This result suggests that during this time the longitudinal distribution of magnetic flux was more organized and not random.

\begin{figure*}[!htbp]
\resizebox{\hsize}{!}{\includegraphics{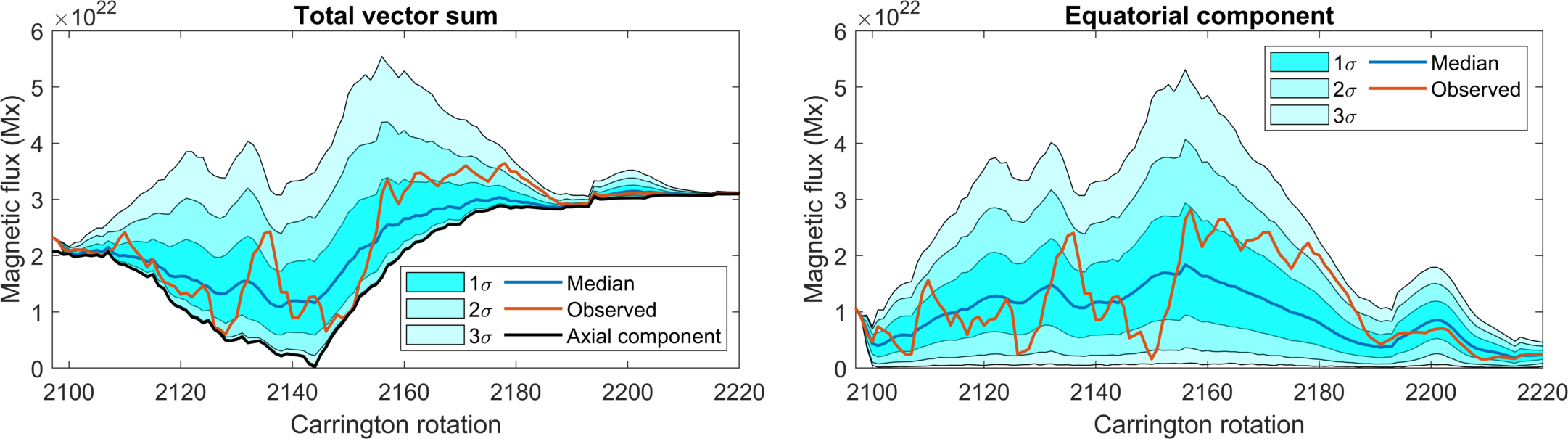}}
\caption{Simulations of large-scale magnetic field with uniformly random longitudes. Left: Vector sum magnitude. Right: Equatorial component of the vector sum. The blue line shows the median dipole strength from $10^4$ simulations. The orange line shows the result from the original simulation. Shaded regions show the 1$\sigma$-3$\sigma$ ranges from the simulations. The black line in the left panel shows the strength of the axial component.
}\label{fig:DipoleFluxRandomLongitudes}
\end{figure*}

Figure~\ref{fig:DipoleFluxRandomLongitudes} shows the strength of the large-scale magnetic field from $10^4$ simulations in which we randomized the emergence longitudes of all active regions.
The left panel shows the total vector sum and the right panel shows the strength of the equatorial component.
Different shadings correspond to 1$\sigma$-3$\sigma$ ranges.
Again, instead of running $10^4$ SFT simulations, we shifted the active region vectors uniformly randomly in longitude and summed the resulting randomized vector ensembles.

Figure~\ref{fig:DipoleFluxRandomLongitudes} shows that different longitude realizations lead to a large variability in large-scale field strength.
The maximum in randomized simulations was reached at CR2156 (October 2014), which shows that timing of the observed peak of the large-scale solar magnetic field at CR2157 (November 2014) occurred around the time when it was most likely.
Before the rapid increase culminating at CR2157 the equatorial component of the large-scale magnetic field almost vanished at CR2150 (May 2014) in simulation with the original active regions.
Only 0.57\% of the simulations with random longitudes reach equally small values at CR2150.
This indicates that the positioning of the real active regions around the solar maximum (CR2149, April 2014, SILSO 13-month smoothed mean) was not random but highly specific with the tendency to cancel the equatorial magnetic field component.
After this minimum value, the equatorial component rapidly increases from the minimum value to the maximum value only in seven rotations.
From the peak onward, the original large-scale strength stays systematically above the median simulation for more than three years until CR2197 (November 2017).

\begin{figure*}[!htbp]
\resizebox{\hsize}{!}{\includegraphics{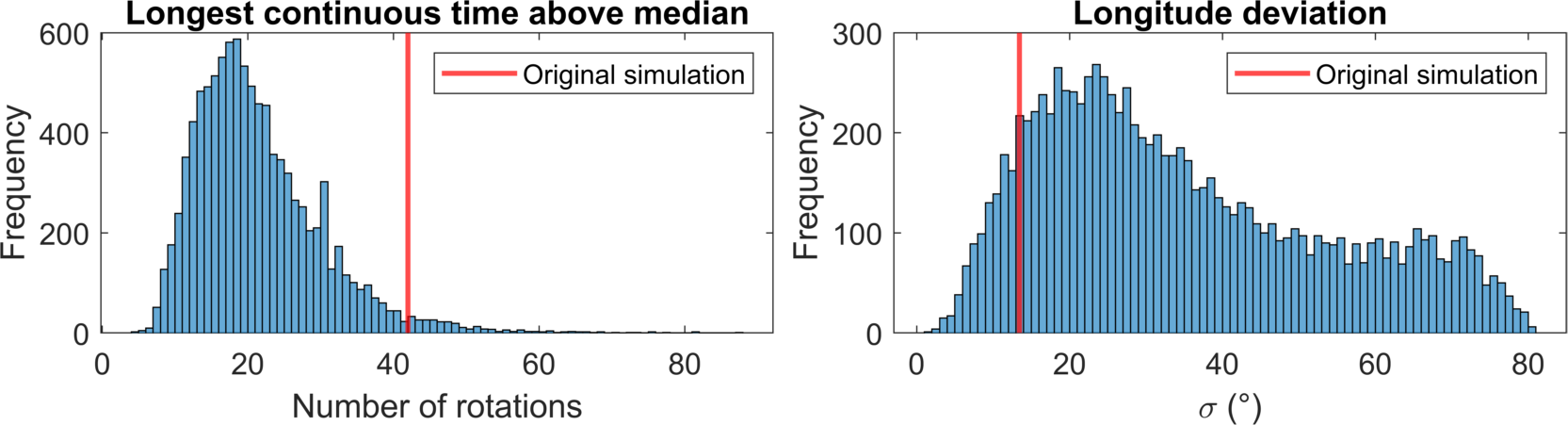}}
\caption{Random longitude statistics. Left panel: Longest continuous time that the strength of the equatorial component was larger than the median of all simulation. Right panel:  Standard deviation of longitude distribution during longest stretch above the median.
}\label{fig:RandomStatistics}
\end{figure*}

Figure~\ref{fig:RandomStatistics} shows two histograms that compare the observed longitude simulation with the random longitude simulations.
The left panel of Fig.~\ref{fig:RandomStatistics} shows the length of the longest period that the equatorial component of the large-scale magnetic field strength stayed above the median level of all simulations (blue line in Fig.\ref{fig:DipoleFluxRandomLongitudes}).
In the simulation with the original longitude distribution, the strength of the large-scale field stayed above this median level for 42 subsequent rotations from CR2155 to CR2197 (September 2014 to November 2017), which is longer than in 97.3\% of the simulations.
Fig.~\ref{fig:RandomStatistics}b shows, for each simulation, the standard deviation of the vector sum longitude during its longest interval above the median of all simulations.
In the original simulation, the standard deviation of the vector sum longitude from CR2155 to CR2197 was 13.4\degree{}, which is smaller than in 89.3\% of the random longitude simulations during their respective above-median intervals.
This indicates that the equatorial direction of the large-scale field was rather stationary during this period.
The stationarity of the vector sum longitude during this period (September 2014 to November 2017) is also evident in Fig.~\ref{fig:VectorSumAndOSF}c.

\section{Discussion}\label{sec:Discussion}
We used the recently developed vector sum method \citep{Tahtinen2024} together with an SFT model to simulate and analyze the evolution of active regions and their effect on the large-scale solar magnetic field during solar cycle 24.
We also studied how changing the longitude distribution of active regions affects the large-scale solar magnetic field.
We exploited the linearity and longitudinal symmetry of the SFT model which, combined with the vector sum, allowed us to represent the evolution of active regions as vectors.
This enabled us to efficiently explore different realizations of longitude distributions without having to run the full simulation for each case.

\subsection{SFT parameters}
The diffusion, $\eta$, and meridional flow amplitude, $u_0$, parameters of the SFT model need to be optimized to match observations.
As our goal was to study the evolution of the large-scale magnetic field, we used the vector sum magnitude as an optimization metric.
We also optimized the SFT model with respect to the axial and equatorial components of the vector sum.
Figure~\ref{fig:BestSFTParameters} shows that the use of the different components leads to different sets of optimal SFT parameters.
Figure~\ref{fig:AxialAndEquatorialParameters} shows that for axial and equatorial components, there is a parameter range that produces almost equally well-performing results, but that these ranges have an opposite behavior.
For the axial component, we find that increasing both components together leads to a set of well-performing solutions, while for the equatorial component, the increase in one parameter is compensated by a decrease in the other.

This behavior is due to the opposite effect that diffusion has on the axial and equatorial components of the large-scale magnetic field, with larger diffusion increasing the axial component but decreasing the equatorial component.
While diffusion alone would eventually cancel both the axial and equatorial components, meridional flow changes the situation so that diffusion actually acts to strengthen the axial component \citep{WangSheeley1991}.
This is because diffusion can transport some of the leading low-latitude magnetic flux of active regions across the equator, causing a hemispheric flux imbalance.
The meridional flow then amplifies the axial component by carrying the leading-polarity flux to the opposite pole.
The stronger meridional flow weakens the axial component because it inhibits the cross-equatorial flux transport.
For the equatorial component, there is no such antisymmetry in the longitudinal direction that the diffusion could amplify.
Therefore, the diffusion works to weaken the equatorial component from the start.
Likewise, the effect of meridional flow is to transport the magnetic flux to the poles, which eventually cancels the equatorial component.
As shown in Fig.~\ref{fig:AxialAndEquatorialParameters}, increasing either of the parameters decreases the equatorial component.

The degeneracy between $\eta$ and $u_0$ for the axial component, where increasing both parameters leads to almost equally good SFT performance within the parameter range, is well-known in the literature \citep[see, e.g.,][]{Yeates2023}.
This degeneracy is related to the dynamo effectivity range $\lambda_R$, which characterizes the latitude at which the axial amplification of the active region dipoles becomes effective \citep{Petrovay2020,Talafa2022}.
Because $\lambda_R$ depends on the ratio $\eta/u_0$, the optimal $\lambda_R$ corresponds to a range of values of $\eta$ and $u_0$, thus explaining the degenerate band in the left panel of Fig.~\ref{fig:AxialAndEquatorialParameters}.
For our optimized parameters $\eta~=~350~\mathrm{km^2/s}$ and $u_0~=~11$~m/s the dynamo effectivity range $\lambda_R=7.4\degree{}$.
This value of $\lambda_R$ agrees with the results of \citet{Yeates2025}, who reconstructed the evolution of photospheric magnetic fields for solar cycles 16 -- 21 using active regions determined from Ca~II~K observations.

We are not aware of the analogous result for the equatorial component shown in the right panel of Fig.~\ref{fig:AxialAndEquatorialParameters}, where increasing $\eta$ and decreasing $u_0$ lead to almost equally good performance for the equatorial component.
In the literature, attention is usually given to the axial component.
For example, the two most general studies \citep{Lemerle2015,Whitbread2017} compared the longitudinally averaged time-latitude butterfly diagrams.
However, the time-latitude maps average over longitudes and destroy the information on the equatorial component.
For example, a zero strength in a butterfly diagram can correspond to both zero and an extremely strong equatorial component, depending on the longitudinal distribution of flux.
Although the best-fit parameters vary between studies, they all find a similar degenerate band that arises from their focus on the axial component.
We find that this degenerate band can be reduced by taking into account the oppositely behaving equatorial component, which essentially selects the best-fit solution out of the degenerate band of the axial component.
Equatorial component, however, does not provide much new information about the dynamo effectivity range as different $\lambda_R$ are already close to each other. 

Recently, \citet{Rincon2025} used HMI observations to characterize turbulent diffusion at unprecedented temporal and spatial scales.
Using two different methods, they estimated turbulent diffusion to be $\eta = 200$–-$400~\mathrm{km^2/s}$ and argued that their results are sufficiently robust to resolve the degeneracy between $\eta$ and $u_0$.
By taking the equatorial component of the solar magnetic field into account, we are likewise able to reduce this degeneracy and obtain a compatible estimate of $\eta~=~350~\mathrm{km^2/s}$.
These results support the use of vector sum (or total solar dipole) as a goodness-of-fit measure, so that both the axial and equatorial components are taken into account.

\subsection{Vector sum and OSF}
The vector sum method \citep{Tahtinen2024} that we use is an efficient way to quantify the strength and orientation of the large-scale dipole of the solar magnetic field.
The vector sum is related to the solar dipole field resulting from the first-order ($l=1)$ spherical harmonic expansion of the magnetic field (see Appendix~\ref{appendix}).
The axial and equatorial components of the vector sum correspond to axial and equatorial dipole moments of the Sun and are related by $|V|=\frac{4\pi R_\odot^2}{3}|D|$, where \textit{V} is the magnitude of the vector sum and \textit{D} the solar dipole moment.

Interestingly, as shown in Figs.~\ref{fig:VectorSumAndOSF} and \ref{fig:VectorSumAndOSFProjections} and discussed in more detail in \citet{Tahtinen2024}, the magnitude of the vector sum closely matches the OSF of the PFSS model with the typically used source surface radius of $R_{ss}=2.5R_\odot{}$.
\citet{Tahtinen2024} also show that, on average, $R_{ss}=2.4-2.5R_\odot{}$ produces the best match between the vector sum and the PFSS OSF on six different magnetogram datasets.

Now, because the vector sum magnitude equals the total magnetic flux aligned with the solar dipole axis (see Appendix~\ref{appendix}), the good match between the vector sum and the PFSS OSF with $R_{ss}=2.5R_\odot{}$ then indicates that the standard choice of the source surface radius corresponds to a distance that produces an OSF that matches the total photospheric flux aligned with the solar dipole in the photosphere.
This observation leads to the question whether the result somehow reflects the structure of the large-scale solar magnetic field.
If we interpret the vector sum magnitude as open flux, then one interpretation of this result is that locally the amount of open flux depends on the dot product between the magnetic field and the unit dipole vector $\mathbf{B_r}\cdot\hat{\mathbf{D}}$.
For a purely radial magnetic field that shares the polarity with the dipole field $\mathbf{B_r}\cdot\hat{\mathbf{D}}>0$, and the magnetic field is entirely open at the dipole's poles and closes down proportionally to the cosine of the angular distance from the poles.
In contrast, regions of opposite polarity relative to the dipole field contribute negatively, as $\mathbf{B_r}\cdot\hat{\mathbf{D}}<0$.
This negative contribution can be interpreted as the closing of otherwise open magnetic field lines.

\subsection{Active regions driving the rapid strengthening of the large-scale magnetic field in 2014}
Since the equatorial component of active regions typically increases for several rotations after their emergence \citep{WangSheeley1991}, the majority of most influential active regions at any given time likely emerged many rotations earlier.
As shown in Fig.~\ref{fig:OSFPeakAREvolution}, this is also true for the peak of the large-scale solar magnetic field that reached its maximum strength at CR2157 (November 2014).
For example, AR\#2 whose equatorial component was the second strongest at CR2157, emerged already 12 rotations earlier at CR2145 (December 2013).

Figure~\ref{fig:ARDipoleLongitude} shows that in the SFT simulation of solar cycle 24, the vector sum longitude of the large-scale field abruptly made a 180\degree{} flip from 60\degree{} to 240\degree{} between CR2149 and CR2150 (April-May 2014) at the time of the solar maximum (SILSO, 13-month smoothed mean).
The right panel of Fig.~\ref{fig:ARDipoleLongitude} shows that the longitude range to which the large-scale field jumps was already building up before CR2150, indicating that the active regions whose equatorial components were aligned with the new direction first weakened the oppositely directed large-scale equatorial component and from CR2150 began to strengthen the new configuration of the large-scale magnetic field.
This new configuration was then further reinforced by the recurrent flux emergence of active regions whose equatorial components were aligned with the large-scale field.
We note that recent analysis of longitudinal activity patterns by \citet{Korsos2025} shows a similar rapid evolution of longitudes in both photospheric and chromospheric magnetic fields as well as in solar flares during this period (see their Fig.~7).

We find that all six of the strongest active regions that contributed constructively to the large-scale magnetic field at the time of the OSF maximum emerged within a narrow longitude band in the southern hemisphere within seven rotations before CR2157 (see our Fig.~\ref{fig:Magnetograms}).
Remarkably, the importance of this longitude band for strengthening the solar equatorial dipole was recognized by \citet{WangSheeley2015} almost immediately.
They reported that the increase resulted from the in-phase emergence of active regions that culminated in the emergence of NOAA AR 12192 at CR2156, which was the largest active region in 24 years.
Here, we arrive at the same conclusion that recurrent flux emergence in this longitude band and its effect of the equatorial component were responsible for the rapid strengthening of large-scale magnetic field in the late 2014.

Figure~\ref{fig:OSFPeakAREvolution} shows that although NOAA AR 12192 was the largest active region in 24 years, its effect on the large-scale field at the OSF maximum was not exceptional compared to other active regions that had had time to grow their equatorial components for multiple rotations. 
For example, if AR\#3 had emerged on the other side of the Sun, it would have decreased the peak equatorial component by 40\%, leading to a more gradual strengthening of the equatorial component (see Fig.\ref{fig:180LongitudeShift}).
On the other hand, if AR\#2 had emerged at the opposite longitude, the rapid strengthening of the large-scale field would have started five rotations earlier, leading to 40\% larger flux at CR2157.

The importance of recurrent flux emergence within the southern longitude band has also been noted more recently.
\citet{Wang2020} show that this longitude band was the main culprit behind the poleward surge that had a significant effect on the evolution of the southern hemisphere during the last half of Solar Cycle 24.
\citet{Finley2024} find that recurrent flux emergence effectively locked the heliospheric current sheet in place for several rotations.
\citet{Heinemann2024} find a potential link between the rapid enhancement of OSF and NOAA AR 12192 that emerged within this longitude band.

\subsection{Longitude distribution of active regions in the descending phase of solar cycle 24}
The strength of the equatorial component of the solar magnetic field remained at a high level for a three-year period following its rapid increase in 2014.
During this period and until the start of 2019, the longitudinal direction of the large-scale magnetic field remained quite stationary in the Carrington frame, as shown in the lower panel of Fig.~\ref{fig:VectorSumAndOSF}.
This period also shows up in Fig.~\ref{fig:PeakFluxSingleAR}.
During this period, the longitudinal configuration of active regions was such that the minimum strength of the equatorial component resulting from shifting a single active region in longitude remained at a fairly high level, contrary to the most the ascending and maximum phase of solar cycle 24, during which the minimum level was mostly close to zero.
This result indicates that during this period the longitude distribution of active regions was highly specific, so that their equatorial components were strengthening each other.

This period also appears in Fig.~\ref{fig:DipoleFluxRandomLongitudes}.
During this period, the strength of the equatorial component is at a higher level than in most of the random longitude simulations. 
The period of elevated equatorial activity begins after the equatorial component first reached almost zero at CR2150.
Only 0.57\% of the randomized simulations obtain a smaller value at CR2150.
The strength of the large-scale field then exceeded the median of randomized simulations at CR2155 (September 2014) and stayed above the median for 42 subsequent rotations until CR2197 (November 2017).
Only 2.7\% of the random simulations contain a longer stretch above the median level (see left panel of Fig.~\ref{fig:RandomStatistics}).
This result indicates that during this period, the active regions had a tendency to appear at longitudes that reinforced the equatorial component of the large-scale field.
The right panel of Fig.~\ref{fig:RandomStatistics} shows that the standard deviation of the vector sum longitude during this period was lower than in 89.3\% of the longest above-median intervals in the random-longitude simulations.
The small longitude deviation indicates that the longitudinal direction of the large-scale field was more stationary than expected from randomly distributed active regions.
The stationarity of the large-scale field during this period agrees with the result that the active regions had a tendency to reinforce the equatorial component of the large-scale field during the declining phase of the solar cycle 24.

\section{Conclusions}\label{sec:Conclusions}
We showed that taking the equatorial component of the solar magnetic field into account can help to better constrain the parameter space of the SFT model than the axial component alone, which is typically emphasized when optimizing SFT models.
The increased discriminative power comes from the fact that the equatorial component has an opposite behavior with respect to the diffusion parameter.

When studying the rapid formation of the maximum of solar cycle 24, we found that recurrent flux emergence that occurred over many rotations around Carrington longitude 240\degree{} in the southern hemisphere played a crucial role in the rapid strengthening of the large-scale equatorial component in 2014.
The equatorial components of these active regions were aligned with the large-scale field, which acted to further strengthen the existing large-scale field.
These results align with the view presented by \citet{WangSheeley2015}.

We found that the longitude distribution of active regions during the latter half of solar cycle 24 was highly specific.
The strength of the large-scale equatorial component was stronger and the longitudinal direction more stationary than expected from a purely uniform distribution during this period. Although our study is focused on solar cycle 24,
the methods presented here can easily be applied to other cycles; for example, to the ongoing solar cycle 25.
Figure~\ref{fig:VectorSumAndOSFProjections} shows that the equatorial component of the large-scale solar magnetic field has already experienced two significant enhancements during the current cycle, with the latter starting in early 2025.
As the equatorial component of an active region typically takes about six rotations to reach its maximum strength, it might be feasible to forecast an enhancement of the equatorial component within this time frame.

\begin{acknowledgements}
I.T. and T.A. acknowledge the financial support by the Research Council of Finland to the SOLEMIP (project no. 357249).
I.T. acknowledges the financial support by the Jenny and Antti Wihuri Foundation.
We thank the anonymous referee for their constructive comments that improved this article.
The authors wish to acknowledge CSC – IT Center for Science, Finland, for computational resources.
\end{acknowledgements}

\bibliography{bibliography} 

\begin{appendix}
\section{Relation of vector sum, solar dipole moment, and PFSS OSF}\label{appendix}
\subsection{Dipole moment}
The Cartesian components of the vector sum $\vec{V}$ can be written as
\begin{align}
    V_x &= R_\odot^2 \int B_r(\theta, \phi) \sin\theta \cos\phi \, d\Omega ,\\
V_y &= R_\odot^2 \int B_r(\theta, \phi) \sin\theta \sin\phi \, d\Omega ,\\
V_z &= R_\odot^2  \int B_r(\theta, \phi) \cos\theta \, d\Omega, 
\end{align}
where $\theta$ and $\phi$ are the colatitude and longitude, $|V|=\Phi=\sqrt{V^2_x+V^2_y+V^2_z}$ and $d\Omega$ the differential solid angle.

The vector sum is closely related to the solar dipole moment which corresponds to the first-order term of spherical harmonic expansion of the photospheric magnetic field.
This can be seen by comparing the above components of the vector sum to the equations of the equatorial $D_{eq}=\sqrt{H^2_1+H^2_2}$ and axial $D_{ax}$ dipole moments 
\begin{align}
    H_1 &= \frac{3}{4\pi} \int B_r(\theta, \phi) \sin\theta \cos\phi \, d\Omega ,\\
H_2 &= \frac{3}{4\pi} \int B_r(\theta, \phi) \sin\theta \sin\phi \, d\Omega ,\\
D_{ax} &= \frac{3}{4\pi} \int B_r(\theta, \phi) \cos\theta \, d\Omega. 
\end{align}
Comparison of the above equations shows that magnitude of the vector sum is directly related to dipole moment by $|V|=\frac{4\pi R_\odot^2}{3}|D|$.

\subsection{Parallel flux}
Vector sum equals the total magnetic flux parallel to the dipole axis.
This is easiest to see in coordinates where the z-axis is aligned with the dipole, and the x and y components vanish.
Parallel flux $\Phi_{||}$ then equals
\begin{align}
\Phi_{||} &= R_\odot^2\int \mathbf{B_r}\cdot\hat{\mathbf{D}}~d\Omega =  R_\odot^2\int B_{r} \cos\theta\, d\Omega = |V|.
\end{align}

\subsection{PFSS OSF}
If only the dipole component ($l=1$) is used in the PFSS expansion \citep[see, e.g.,][]{Wang1992} then in coordinates where the dipole is aligned with the z-axis, the (radial) magnetic field at the source surface has form
\begin{align}
B^{ss}_r(\theta) &= C_{rl} \cos\theta \frac{3}{4\pi}  \int B_r \cos\theta d\Omega \\
  &= C_{rl} D\cos\theta.
\end{align}
$C_{rl}$ is a radial function depicting the decay of the magnetic field with altitude.
When $l=1$ it has the following form at the source surface where $r=r_{ss}=\frac{R_{ss}}{R_\odot}$
\begin{equation}
  C_{rl} =\frac{3 }{2r^3_{ss}+1}.
\end{equation}
Integrating $|B^{ss}_r|$ over the source surface gives OSF
\begin{align}
OSF &= C_{rl}R^2_{ss}|D|  \int|\cos\theta| d\Omega  ,\\
&= C_{rl}2\pi R^2_{ss}|D| ,\\
&= \frac{3C_{rl} r^2_{ss} |V|}{2}. \label{eq:OSFVS}
\end{align}
We can solve the source surface radius $r_{ss}$  that corresponds to the vector sum from Eq.~\ref{eq:OSFVS} by setting OSF equal to the vector sum yielding the following equation
\begin{align}
4r^3_{ss}-9r^2_{ss}+2 = 0.
\end{align}
Solving this equation gives $r_{ss}\approx2.141$, that is, the magnitude of the vector sum equals the PFSS OSF when $l_{max}=1$ and $R_{ss}~=~2.141R_\odot$.

\end{appendix}

\end{document}